\documentclass[epsfig,12pt]{article}
\usepackage{epsfig}
\usepackage{graphicx}
\usepackage{hyperref}

\usepackage{array}
\usepackage{amsmath}
\usepackage{amssymb}

\newcommand{\beq}{\begin{equation}}   
\newcommand{\eeq}{\end{equation}}
\newcommand{\beqn}{\begin{eqnarray}}   
\newcommand{\eeqn}{\end{eqnarray}}

\begin{document}
\unitlength = 1mm

\def\de{\partial}
\def\Tr{ \hbox{\rm Tr}}
\def\const{\hbox {\rm const.}}  
\def\o{\over}
\def\im{\hbox{\rm Im}}
\def\re{\hbox{\rm Re}}
\def\bra{\langle}\def\ket{\rangle}
\def\Arg{\hbox {\rm Arg}}
\def\Re{\hbox {\rm Re}}
\def\Im{\hbox {\rm Im}}
\def\diag{\hbox{\rm diag}}

%%%%%%%%%%%%%%%%%%%%%%%%%%%%%%%%%%%%%%%%%%%%%%%%%%%%%%%%%%%%%%%%%%%%

\def\QATOPD#1#2#3#4{{#3 \atopwithdelims#1#2 #4}}
\def\stackunder#1#2{\mathrel{\mathop{#2}\limits_{#1}}}
\def\stackreb#1#2{\mathrel{\mathop{#2}\limits_{#1}}}
\def\Tr{{\rm Tr}}
\def\res{{\rm res}}
\def\Bf#1{\mbox{\boldmath $#1$}}
\def\balpha{{\Bf\alpha}}
\def\bbeta{{\Bf\beta}}
\def\bgamma{{\Bf\gamma}}
\def\bnu{{\Bf\nu}}
\def\bmu{{\Bf\mu}}
\def\bphi{{\Bf\phi}}
\def\bPhi{{\Bf\Phi}}
\def\bomega{{\Bf\omega}}
\def\blambda{{\Bf\lambda}}
\def\brho{{\Bf\rho}}
\def\bsigma{{\bfit\sigma}}
\def\bxi{{\Bf\xi}}
\def\bbeta{{\Bf\eta}}
\def\d{\partial}
\def\der#1#2{\frac{\d{#1}}{\d{#2}}}
\def\Im{{\rm Im}}
\def\Re{{\rm Re}}
\def\rank{{\rm rank}}
\def\diag{{\rm diag}}
\def\2{{1\over 2}}
\def\ntwo{${\mathcal N}=2\;$}
\def\nfour{${\mathcal N}=4\;$}
\def\none{${\mathcal N}=1\;$}
\def\ntwot{${\mathcal N}=(2,2)\;$}
\def\ntwoo{${\mathcal N}=(0,2)\;$}
\def\x{\stackrel{\otimes}{,}}

\newcommand{\cpn}{CP$(N-1)\;$}
\newcommand{\wcpn}{wCP$_{N,\widetilde{N}}(N_f-1)\;$}
\newcommand{\wcpd}{wCP$_{\widetilde{N},N}(N_f-1)\;$}
\newcommand{\wcpt}{$\mathbb{WCP}(2,2)\;$}
\newcommand{\wcpo}{$\mathbb{WCP}(1,1)\;$}
\newcommand{\wcp}{$\mathbb{WCP}(N,\tilde N)\;$}
\newcommand{\vp}{\varphi}
\newcommand{\pt}{\partial}
\newcommand{\tN}{\widetilde{N}}
\newcommand{\ve}{\varepsilon}
\renewcommand{\theequation}{\thesection.\arabic{equation}}

\newcommand{\sun}{SU$(N)\;$}

\setcounter{footnote}0
%\begin{titlepage}
%\renewcommand{\thefootnote}{\fnsymbol{footnote}}

\vfill

%%%%%%%%%%%%%%%%%%%%%%%%%%%%%%%%
\begin{titlepage}

\begin{flushright}
%FTPI-MINN-17/?, UMN-TH-????/17\\
% April 28, 2016
\end{flushright}

\begin{center}
{  \Large \bf  
  Flux Compactification for the Critical Non-Abelian Vortex 
%\\[2mm] 
and  Quark Masses
}

\vspace{5mm}
%\vspace{1mm}

{\large  \bf A.~Yung$^{\,\,a,b}$}
\end {center}

\begin{center}

$^{a}${\it National Research Center ``Kurchatov Institute'', 
Petersburg Nuclear Physics Institute, Gatchina, St. Petersburg
188300, Russia}\\
$^b${\it  William I. Fine Theoretical Physics Institute,
University of Minnesota,
Minneapolis, MN 55455}\\

\end{center}

\vspace{1cm}

\begin{center}
{\large\bf Abstract}
\end{center}

It has been  shown that  non-Abelian solitonic vortex strings supported in four-dimensional (4D)
\ntwo supersymmetric QCD (SQCD) with the U($N=2$) gauge group
and $N_f=4$  quark flavors  behave as critical superstrings. In addition to four
translational moduli non-Abelian strings under consideration carry six orientational and size moduli. 
Together they form a ten-dimensional space required for a superstring to 
be critical.  The  target space of the string sigma model is a product of the flat four-dimensional
space $\mathbb{R}^4$ and a Calabi-Yau non-compact threefold $Y_6$, namely, the conifold. The
spectrum of low lying closed string states in the associated type IIA string theory was found and interpreted
as a spectrum of hadrons in 4D \ntwo  SQCD. In particular,
the lowest string state appears to be a massless BPS baryon associated with the deformation of the complex structure modulus
$b$ of the conifold. Here we address a problem of switching on  quark masses in 4D  SQCD, which classically
breaks the world sheet conformal invariance in the string sigma model. To avoid this problem we follow a standard string theory approach and use a flux ''compactification'' to lift the complex structure modulus of the conifold. Namely, we find a solution of supergravity equations of motion with non-zero   NS 3-form flux. It produces a potential for the 
 baryon $b$, which leads to the  run-away vacuum. Using field theory arguments we interpret 3-form flux in terms of a particular choice  of quark masses in 4D SQCD. At the run-away vacuum
the conifold degenerates to  lower dimensions. We interpret this as a flow from a non-Abelian string to an Abelian one.

\vspace{2cm}

\end{titlepage}

\newpage

%\tableofcontents

\newpage

\section {Introduction }
\label{intro}
\setcounter{equation}{0}

It was shown in \cite{SYcstring} that the non-Abelian solitonic vortex string in  
 4D \ntwo supersymmetric QCD (SQCD) with the U(N=2) gauge group and  $N_f=4$   flavors of quark hypermultiplets becomes a critical  superstring.
Non-Abelian vortices were first found in 
\ntwo  SQCD with the gauge group U$(N)$ and $N_f \ge N$ flavors of quarks
\cite{HT1,ABEKY,SYmon,HT2}. The non-Abelian vortex string is 1/2
BPS saturated and, therefore,  has \ntwot supersymmetry on its world sheet.
In addition to four translational moduli characteristic of the  Abrikosov-Nielsen-Olesen (ANO) strings 
\cite{ANO}, the non-Abelian string carries orientational  moduli, as well as the size moduli if $N_f>N$
\cite{HT1,ABEKY,SYmon,HT2} (see \cite{Trev,Jrev,SYrev,Trev2} for reviews). Their dynamics
are described by the effective two-dimensional (2D) sigma model on the string world sheet with 
the target space 
\beq
\mathcal{O}(-1)^{\oplus(N_f-N)}_{\mathbb{CP}^1}\,,
\label{12}
\eeq
to which we will refer to as to the weighted CP  model ($\mathbb{WCP}(N,N_f-N)$). 

 For $N_f=2N$
the world sheet sigma model becomes conformal. Moreover, for $N=2$ the 
number of the orientational/size moduli  is six and they can be combined with 
four translational moduli to form a ten-dimensional (10D) space required for a
superstring to become critical \cite{SYcstring,KSYconifold}. In this case the target space of the world sheet 2D theory on 
the non-Abelian vortex string is
 $\mathbb{R}^4\times Y_6$, where $Y_6$ is a non-compact six dimensional Calabi-Yau (CY) manifold, the conifold \cite{Candel,NVafa}.

Non-Abelian vortices in  U$(N)$ theories 
are topologically stable and cannot be broken. In particular, monopoles confined by the non-Abelian string are junctions of two strings with different orientations rather then string endpoints, see review \cite{SYrev}. Therefore
the finite-length strings are closed.

 Thus, we focus on the 
closed strings and consider CY ''compactification'' (Kaluza-Klein reduction) of 10D string theory associated with the non-Abelian vortex to 4D on the conifold $Y_6$ \footnote{Here and below we use the term ''compactification'' in quotation marks because the conifold is a non-compact CY space.} . The goal is to  identify the closed string states   with the hadrons of  4D \ntwo SQCD. The vortex string theory at hand was identified as the type IIA string theory \cite{KSYconifold}.

The first step of this program, namely,  identifying massless
string states was carried out in \cite{KSYconifold,KSYcstring} using supergravity approximation.
It turns out that  most of massless modes have  non-normalizable wave functions over the non-compact conifold $Y_6$, i.e. they are not localized in 4D 
and, hence, cannot be interpreted as dynamical states in 4D SQCD. In particular, the 4D graviton and  unwanted vector multiplet associated with deformations of the K\"ahler form of the conifold are absent.
 However, a single massless BPS hypermultiplet   was found 
at the  self-dual point at strong coupling. It is associated with deformations of a complex structure of the conifold and was  interpreted  as a composite 4D baryon $b$ \footnote{ The definition of the baryonic charge is non-standard and will be given below in Sec. \ref{sec:NAstring}.}.

Later  low lying massive non-BPS 4D states were found in \cite{SYlittles,SYlittmult} using the little string theory 
approach, see \cite{Kutasov} for a review.

In this paper we consider 4D \ntwo SQCD with non-zero  quark masses. Quark masses are the only deformations of the 4D
SQCD with  given gauge and flavor groups allowed by \ntwo supersymmetry, which do not involve higher derivative operators.  It is interesting to study
the string theory response to these deformations. However, making quarks massive lifts some of orientational and size moduli
of the non-Abelian string. Generically, this  breaks the  conformal invariance of the world sheet \wcpt model (at least classically)
preventing the formulation of the string theory in this case. This problem is puzzling because  from  the standpoint of 4D
SQCD switching on   quark masses is physically the most natural deformation one can consider.

To overcome this difficulty instead of attempting to interpret mass terms for orientational/size moduli in \wcpt model in terms of 10D supergravity we propose from very beginning to use a flux ''compactification'' to describe 4D quark masses in the string theory framework. The motivation is that fluxes generically induce a potential for CY moduli lifting flat directions, see, for example, \cite{Louis} for a review. On the other hand, in 4D theory we  expect that quark masses also induce a potential
for  the massless  baryon $b$ associated with the conifold complex structure modulus. In particular, we know that it  acquires the mass
\beq
m_{b} = |m_1+m_2-m_3-m_4|
\label{m_b}
\eeq
at non-zero $m_A$ dictated for a BPS state by its baryonic charge \cite{ISY_b_baryon}, where $m_A$, $A=1,...,4$ are quark masses. Thus, we  expect that fluxes in 10D supergravity  can be  interpreted  in terms of quark masses of 4D SQCD.

Generically, fluxes modify the metric so that the deformed background is a solution of 10D supergravity equations of motion. This  guaranties  the conformal invariance of the deformed world sheet sigma model.

It is known that for type IIA CY compactifications the potential for the  K\"ahler form moduli arise from RR even-form fluxes,
while  the potential for complex structure moduli  is induced by NS 3-form flux
$H_3$  \cite{Louis2,Kachru}. Since we are interested in the potential for the conifold complex structure modulus $b$
we consider the NS 3-form $H_3$. It does not break \ntwo supersymmetry in 4D theory \cite{Louis2}.

 We use a perturbation theory to solve 10D gravity equations solving first equations of motion for $H_3$ using the conifold metric.  The back reaction on the metric and the dilaton arise in the quadratic order in $H_3$ and we neglect this effect to the leading order at small $H_3$. 

Note that if we keep the vacuum expectation value (VEV) of the baryon $b$ large the curvature of the conifold is everywhere small and the gravity  approximation is justified.  

It turns out that $H_3$ flux does not generate a mass term \eqref{m_b} for the baryon $b$. Mass term vanishes due to the non-compactness of the conifold. Instead, $H_3$ flux produces a potential leading to a run-away vacuum for $b$. We still  look for the interpretation of  $H_3$ flux in terms of quark masses. The reason for this is that there is no
other deformation in 4D SQCD ( preserving \ntwo supersymmetry), which produces a scalar potential except mass deformation. 
Using field theory arguments we relate $H_3$ to of a particular choice  of quark masses in 4D SQCD. 
At the run-away vacuum
the deformed conifold degenerates to  lower dimensions. We interpret this as a flow from a non-Abelian string to an Abelian one.
%We interpret this as a flow of the \ntwo SQCD to a theory with smaller gauge group and quark content in 4D, namely to the Abelian U(1) gauge theory with $N_f=2$ quark flavors.

The paper is organized as follows. In Sec. \ref{sec:NAstring} we briefly review 4D \ntwo SQCD and  the world sheet sigma  model
on the non-Abelian string. Next we review massless baryon $b$ as a deformation of the complex structure of the conifold and  present the conifold metric.  In Sec. \ref{sec:H_3} we solve
equations of motion for the 3-form $H_3$ and show  that $H_3$ does not induce a mass \eqref{m_b} for $b$-baryon. Also 
we derive the potential for $b$ leading to the run-away vacuum. To understand better the behavior of the potential at large $b$ in Sec. \ref{sec:H_3large_b} we solve the equations of motion for $H_3$ using deformed conifold metric and calculate the potential at large $b$.
In Sec. \ref{sec:quarkmasses} we interpret $H_3$-form in terms of quark masses in 4D SQCD. We also discuss the interpretation of the degeneration of the conifold at the run-away vacuum as a flow from a non-Abelian string to an Abelian one. Sec. \ref{sec:conclusions} summarizes  our conclusions.

\section {Non-Abelian critical vortex string}
\label{sec:NAstring}
\setcounter{equation}{0}

\subsection{Four-dimensional \boldmath{${\mathcal N}=2\;$} 
 SQCD}
\label{sec:SQCD}

As was already mentioned, non-Abelian vortex-strings were first found in 4D
\ntwo SQCD with the gauge group U$(N)$ and $N_f \ge N$ quark flavors 
supplemented by the Fayet-Iliopoulos (FI) \cite{FI} term with parameter $\xi$
\cite{HT1,ABEKY,SYmon,HT2}, see for example, \cite{SYrev} for a detailed review of this theory.
Here, we just mention that at weak coupling $g^2\ll 1$, this theory is in the Higgs phase in which the scalar
components of the quark multiplets (squarks) develop vacuum expectation values (VEVs). These VEVs break 
the U$(N)$ gauge group
Higgsing  all gauge bosons. The Higgsed gauge bosons combine with the screened quarks to form long \ntwo multiplets with mass $m_G \sim g\sqrt{\xi}$.

 The global flavor SU$(N_f)$ is broken down to the so called color-flavor
locked group. The resulting global symmetry is
\beq
 {\rm SU}(N)_{C+F}\times {\rm SU}(N_f-N)\times {\rm U}(1)_B,
\label{c+f}
\eeq
see \cite{SYrev} for more details. 

The unbroken global U(1)$_B$ factor above is identified with a baryonic symmetry. Note that 
what is usually identified as the baryonic U(1) charge is a part of  our 4D theory  gauge group.
 ``Our" U(1)$_B$
is  an unbroken by squark VEVs combination of two U(1) symmetries;  the first is a subgroup of the flavor 
SU$(N_f)$, and the second is the global U(1) subgroup of U$(N)$ gauge symmetry.

As was already noted, we consider \ntwo SQCD  in the Higgs phase:  $N$ squarks  condense. Therefore,  non-Abelian 
vortex strings confine monopoles. In the \ntwo 4D theory these strings are 1/2 BPS-saturated; hence,  their
tension  is determined  exactly by the FI parameter,
\beq
T=2\pi \xi\,.
\label{ten}
\eeq
However, as we already mentioned,
the monopoles cannot be attached to the string end points. In fact, in the U$(N)$ theories confined  
 monopoles 
are  junctions of two distinct elementary non-Abelian strings \cite{T,SYmon,HT2} (see \cite{SYrev} 
for a review). As a result,
in  four-dimensional \ntwo SQCD we have 
monopole-anti-monopole mesons in which the monopole and anti-monopole are connected by two confining strings.
 In addition, in the U$(N)$  gauge theory we can have baryons  appearing as  a closed 
``necklace'' configurations of $N\times$(integer) monopoles \cite{SYrev}. For the U(2) gauge group the 
important example of a baryon consists of four monopoles \cite{ISY_b_baryon}.

Below we focus on the particular case $N=2$ and $N_f=4$ because, as was mentioned in the Introduction, in this case 4D \ntwo SQCD supports non-Abelian vortex strings which behave as critical superstrings \cite{SYcstring}.  Also, for $N_f=2N$ the gauge coupling $g^2$ of the 4D SQCD does not run; the $\beta$ function 
vanishes. However, the conformal invariance of the 4D theory is explicitly broken by the FI parameter $\xi$, which 
defines VEV's of quarks. The FI parameter is not renormalized.

Both stringy monopole-antimonopole mesons and monopole baryons with spins $J\sim 1$ have masses determined 
by the string tension,  $\sim \sqrt{\xi}$ and are heavier at weak coupling $g^2\ll 1$ than perturbative states with masses
$m_G\sim g\sqrt{\xi}$. 
Thus, they can decay into perturbative states \footnote{Their quantum numbers with respect to the global group 
\eqref{c+f} allow these decays, see \cite{SYrev}.} and in fact at weak coupling we do not 
expect them to appear as stable  states.

Only in the   strong coupling domain $g^2\sim 1$  we expect that (at least some of) stringy mesons and baryons become stable.
These expectations were confirmed in \cite{KSYconifold,SYlittles} where low lying string states in the string theory for the critical non-Abelian vortex were found at the self-dual point at strong coupling.

Below in this paper we introduce small quark masses $m_A$, $A=1,...4$ assuming that two first squark flavors with masses $m_1$ and $m_2$ develop VEVs.

 %If we introduce non-zero quark masses the Higgs branch \eqref{dimH} is lifted and 
%bifundamental quarks acquire masses $(m_P-m_K)$, $P=1,2$, $K=3,4$. Note that bifundamental quarks form  short BPS multiplets
%and their masses do not receive quantum corrections, see \cite{SYrev} for details.

\subsection{World-sheet sigma model}
\label{sec:wcp}

The presence of the color-flavor locked group SU$(N)_{C+F}$ is the reason for the formation of 
non-Abelian vortex strings \cite{HT1,ABEKY,SYmon,HT2}.
The most important feature of these vortices is the presence of the  orientational  zero modes.
As was already mentioned, in \ntwo SQCD these strings are 1/2 BPS saturated. 

Let us briefly review the model emerging on the world sheet
of the non-Abelian  string \cite{SYrev}.

The translational moduli fields  are described by the Nambu–-Goto action and  decouple from all other moduli. Below we focus on
 internal moduli.

If $N_f=N$  the dynamics of the orientational zero modes of the non-Abelian vortex, which become 
orientational moduli fields 
 on the world sheet, are described by 2D
\ntwot supersymmetric ${\mathbb{CP}}(N-1)$ model.

If one adds additional quark flavors, non-Abelian vortices become semilocal --
they acquire size moduli \cite{AchVas}.  
In particular, for the non-Abelian semilocal vortex in U(2) \ntwo SQCD with four flavors,  in 
addition to  the complex orientational moduli  $n^P$ (here $P=1,2$), we must add two complex size moduli   
$\rho^K$ (where $K=3,4$), see \cite{HT2,HT1,AchVas,SYsem,Jsem,SVY}. 

The effective theory on the string world sheet is a two-dimensional \ntwot supersymmetric \wcpt model
 \footnote{Both the orientational and the size moduli
have logarithmically divergent norms, see e.g.  \cite{SYsem}. After an appropriate infrared 
regularization, logarithmically divergent norms  can be absorbed into the definition of 
relevant two-dimensional fields  \cite{SYsem}.
In fact, the world-sheet theory on the semilocal non-Abelian string is 
not exactly the \wcp  model \cite{SVY}, there are minor differences. The actual theory is called the $zn$ model. Nevertheless it has the same infrared physics as the model (\ref{wcp22}) \cite{KSVY}, see also \cite{CSSTY}.} %
\cite{SYcstring,KSYconifold,KSYcstring}%
.
This model describes internal dynamics of the non-Abelian semilocal string. 
For details see e.g. the review \cite{SYrev}.

The \wcpt sigma model 
can be  defined  as a low energy limit of the  U(1) gauge theory \cite{W93}, which corresponds to taking the limit 
of infinite gauge coupling, $e^2\to\infty$ in the action below.  The bosonic part of the action reads
 \footnote{Equation 
(\ref{wcp22}) and similar expressions below are given in Euclidean notation.}
\begin{equation}
\begin{aligned}
	&S = \int d^2 x \left\{
	\left|\nabla_{\alpha} n^{P}\right|^2 
	+\left|\tilde{\nabla}_{\alpha} \rho^K\right|^2
	+\frac1{4e^2}F^2_{\alpha\beta} + \frac1{e^2}\,
	\left|\pt_{\alpha}\sigma\right|^2
	\right.
	\\[3mm]
	&+\left.
	2\left|\sigma+\frac{m_P}{\sqrt{2}}\right|^2 \left|n^{P}\right|^2 
	+ 2\left|\sigma+\frac{m_{K}}{\sqrt{2}}\right|^2\left|\rho^K\right|^2
	+ \frac{e^2}{2} \left(|n^{P}|^2-|\rho^K|^2 - {\rm Re}\,\beta \right)^2
	\right\},
	\\[4mm]
	&
	P=1,2\,,\qquad K=3,4\,.
\end{aligned}
\label{wcp22}
\end{equation}
Here, $m_A$ ($A=1,..,4$) are the so-called twisted masses (they coincide with  4D quark masses),
while ${\rm Re}\,\beta$ is the inverse coupling constant (2D FI term). More  exactly it is the real part of the complexified coupling constant introduced in Eq. (\ref{beta_complexified}) below.

The fields $n^{P}$ and $\rho^K$ have
charges  $+1$ and $-1$ with respect to the auxiliary U(1) gauge field, and the corresponding  covariant derivatives in (\ref{wcp22}) are defined as 
\begin{equation}
	\nabla_{\alpha}=\pt_{\alpha}-iA_{\alpha}\,,
	\qquad 
	\tilde{\nabla}_{\alpha}=\pt_{\alpha}+iA_{\alpha}\,,	
\end{equation}
respectively. The complex scalar field $\sigma$ is a superpartner of the U(1) gauge field $A_{\alpha}$.

Apart from the U(1) gauge symmetry, the sigma model (\ref{wcp22}) in the massless limit has a global symmetry group
\begin{equation}
	 {\rm SU}(2)\times {\rm SU}(2)\times {\rm U}(1)_B \,,
\label{globgroup}
\end{equation}
i.e. exactly the same as the unbroken global group in the 4D theory at $N=2$ and $N_f=4$. 
The fields $n$ and $\rho$ 
transform in the following representations:
\begin{equation}
	n:\quad \left(\textbf{2},\,\textbf{1},\, 0\right), \qquad \rho:\quad \left(\textbf{1},\,\textbf{2},\, 1\right)\,.
\label{repsnrho}
\end{equation}
Here  the global ``baryonic''  U(1)$_B$ symmetry is a classically unbroken (at $\beta >0$) combination of the 
global U(1) group which
rotates $n$ and $\rho$ fields with the same phases plus  U(1) gauge symmetry which  rotates them with the opposite phases, see
\cite{KSYconifold} for details.
Non-zero twisted masses $m_A$ break each of the SU(2) factors in \eqref{globgroup} down to U(1).

The 2D coupling constant ${\rm Re}\,\beta$ can be naturally complexified if we
include the $\theta$ term in the action,
\begin{equation}
	\beta = {\rm Re}\,\beta + i \, \frac{\theta_{2D}}{2 \pi} \,,
\label{beta_complexified}	
\end{equation}
where $\theta_{2D}$ is the 2D $\theta$ angle. 

The number of real bosonic degrees of freedom in the model \eqref{wcp22} is $8-1-1=6$.  Here 8 is the number of real degrees of 
freedom of $n^P$ and $\rho^K$ fields and we subtracted one real constraint imposed by the the $D$ term condition in \eqref{wcp22} 
\beq
|n^{P}|^2-|\rho^K|^2 = {\rm Re}\,\beta,
\label{D-term}
\eeq
 in the limit $e^2\to \infty$  and one  gauge phase eaten by the Higgs mechanism. As we already mentioned, these six internal degrees of freedom in the massless limit can be combined with four translational moduli to form a 10D space needed for a superstring to be critical.

At  the quantum level, the coupling $\beta$ does not run in this theory. Thus, 
the \wcpt  model is superconformal at zero masses $m_A = 0$. Therefore, its target space is Ricci flat and  (being K\"ahler due to \ntwot supersymmetry) represents  a non-compact Calabi-Yau manifold,  namely the conifold $Y_6$, see \cite{NVafa} for a review.

The \wcpt model \eqref{wcp22} with $m_A=0$ was used in \cite{SYcstring,KSYconifold} to define 
the critical string theory for the non-Abelian vortex at hand.

Typically solitonic strings are ''thick'' and the effective world sheet theory like the one in \eqref{wcp22} has 
a series of unknown high derivative corrections in powers of  $\pt/m_G$.
The string transverse size is given  by $1/m_G$, where $m_G \sim g\sqrt{\xi}$ is 
a  mass scale  of the gauge bosons and quarks  forming the string. The string cannot be thin in a  weakly
coupled 4D SQCD because at weak coupling $m_G\sim g\sqrt{T}$ and $m_G^2$ is always small in the units of the string tension $T$,
see \eqref{ten}.

A conjecture was put forward in  \cite{SYcstring}  that  at strong coupling 
in the vicinity of a critical value  $g_c^2\sim 1$ the non-Abelian string in the theory at hand  becomes thin,
and higher-derivative corrections in the world sheet theory \eqref{wcp22} are absent. This is possible because the low energy sigma model \eqref{wcp22} already describes a critical string and higher-derivative corrections are not required to improve its 
ultra-violet behavior, see \cite{PolchStrom} for the discussion of this problem.
 The above  conjecture implies that $m_G(g^2) \to \infty$ at  $ g^2\to g_c^2$. As expected 
the thin string produces linear Regge trajectories even for small spins \cite{SYlittmult}.
 
It was also conjectured in \cite{KSYconifold} that $g_c$ corresponds to the value of the 2D coupling constant $\beta=0$.
The motivation for this conjecture is that  this value is a self-dual point for the \wcpt model \eqref{wcp22}. Also $\beta=0$ is a natural choice because at this point we have a regime change in the \wcpt model. The resolved conifold defined by the $D$ term condition \eqref{D-term} develops a conical  singularity at this point.  The point $\beta=0$ corresponds to 
$\tau_{SW} =1$ in the 4D SQCD, where $\tau_{SW}$ is the complexified inverse coupling, $\tau_{SW}= i\frac{8\pi}{g^2} 
+ \frac{\theta_{4D}}{\pi}$, where $\theta_{4D}$ is the 4D $\theta$ angle \cite{ISY_b_baryon}.

A version of the string-gauge duality
for 4D SQCD  was proposed \cite{SYcstring}: at weak coupling this 
theory is in the Higgs phase and can be 
described in terms of quarks and Higgsed gauge bosons, while at strong coupling hadrons of this theory 
can be understood as string states formed by the non-Abelian vortex string.

At non-zero quark masses $m_A\neq 0$ the model \eqref{wcp22} is a mass deformation of the  superconformal CY theory
on the conifold. Generically  quark masses break the world sheet conformal invariance. The  \wcpt model in \eqref{wcp22} can no longer be used to define a string theory for the non-Abelian vortex in the massive 4D SQCD.

\subsection {Massless 4D baryon}
\label{conifold}
 %\setcounter{equation}{0}

%Before turning on quark masses
In this section we briefly review the only 4D massless state found in the string theory of the critical non-Abelian vortex
in the massless limit \cite{KSYconifold}. It is associated 
with the deformation of the conifold complex structure. 
 As was already mentioned, all other massless string modes  have non-normalizable wave functions over the conifold. In particular, 4D graviton associated with a constant wave
function over the conifold $Y_6$ is
absent \cite{KSYconifold}. This result matches our expectations since we started with
\ntwo SQCD in the flat four-dimensional space without gravity.

We can construct the U(1) gauge-invariant ``mesonic'' variables
\beq
w^{PK}= n^P \rho^K.
\label{w}
\eeq
These variables are subject to the constraint
\beq
{\rm det}\, w^{PK} =0. 
\label{coni}
\eeq

Equation (\ref{coni}) defines the conifold $Y_6$.  
It has the K\"ahler Ricci-flat metric and represents a non-compact
 Calabi-Yau manifold \cite{Candel,NVafa,W93}. It is a cone which can be parametrized 
by the non-compact radial coordinate 
\beq
\widetilde{r}^{\, 2} = {\rm Tr}\, \bar{w}w\,
\label{tilder}
\eeq
and five angles, see \cite{Candel}. Its section at fixed $\widetilde{r}$ is $S_2\times S_3$.

At $\beta =0$ the conifold develops a conical singularity, so both spheres $S_2$ and $S_3$  
can shrink to zero.
The conifold singularity can be smoothed out
in two distinct ways: by deforming the K\"ahler form or by  deforming the 
complex structure. The first option is called the resolved conifold and amounts to keeping
a non-zero value of $\beta$ in (\ref{D-term}). This resolution preserves 
the K\"ahler structure and Ricci-flatness of the metric. 
If we put $\rho^K=0$ in (\ref{wcp22}) we get the $\mathbb{CP}(1)$ model with the sphere $S_2$ as a target space
(with the radius $\sqrt{\beta}$).  
The resolved conifold has no normalizable zero modes. 
In particular, 
the modulus $\beta$  which becomes a scalar field in four dimensions
 has non-normalizable wave function over the 
$Y_6$ and therefore is not dynamical \cite{KSYconifold}.  

If $\beta=0$ another option exists, namely a deformation 
of the complex structure \cite{NVafa}. 
It   preserves the
K\"ahler  structure and Ricci-flatness  of the conifold and is 
usually referred to as the {\em deformed conifold}. 
It  is defined by deformation of Eq.~(\ref{coni}), namely,   
\beq
 {\rm det}\, w^{PK} = b\,,
\label{deformedconi}
\eeq
where $b$ is a complex parameter.
Now  the sphere $S_3$ can not shrink to zero, its minimal size is determined by $b$. 

The modulus $b$ becomes a 4D complex scalar field. The  effective action for  this field was calculated in \cite{KSYconifold}
using the explicit metric on the deformed conifold  \cite{Candel,Ohta,KlebStrass},
\beq
S_{{\rm kin}}(b) = T\int d^4x |\pt_{\mu} b|^2 \,
\log{\frac{\widetilde{R}_{\rm IR}^2}{|b|}}\,,
\label{Sb}
\eeq
where $\widetilde{R}_{\rm IR}$ is the  maximal value of the radial coordinate $\widetilde{r}$  introduced as an infrared regularization of the 
logarithmically divergent $b$-field  norm. Here the logarithmic integral at small $\widetilde{r}$ is cut off by the minimal size of $S_3$, which is equal to $|b|$.

We see that the norm of
the  modulus $b$ turns out to be  logarithmically divergent in the infrared.
The modes with the logarithmically divergent norm are at the borderline between normalizable 
and non-normalizable modes. Usually
such states are considered as ``localized'' ones. We follow this rule. 
This scalar mode is localized near the conifold singularity  in the same sense as the orientational 
and size zero modes are localized on the vortex-string solution.
   
 The field $b$  being massless can develop a VEV. Thus, 
we have a new Higgs branch in 4D \ntwo SQCD which is developed only for the critical value of 
the 4D coupling constant $\tau_{SW}=1$ associated with $\beta=0$.

 In \cite{KSYconifold} the massless state $b$ was interpreted as a baryon of 4D \ntwo QCD.
Let us explain this.
 From Eq.~(\ref{deformedconi}) we see that the complex 
parameter $b$ (which is promoted to a 4D scalar field) is a singlet with respect to both SU(2) factors in
 (\ref{globgroup}), i.e. 
the global world-sheet group.\footnote{Which is isomorphic to the 4D
global group \eqref{c+f} for $N=2$, $N_f=4$.} What about its baryonic charge? From \eqref{repsnrho} and \eqref{deformedconi}
we see that the $b$ state transforms as 
\beq
({\bf 1},\,{\bf 1},\,2).
\label{brep}
\eeq
 In particular it has the baryon charge $Q_B(b)=2$.

 In type IIA superstring compactifications the complex scalar 
associated with deformations of the complex structure of the Calabi-Yau
space enters as a 4D \ntwo BPS hypermultiplet, see \cite{Louis} for a review. Other components of this hypermultiplet can be restored 
by \ntwo supersymmetry. In particular, 4D \ntwo hypermultiplet should contain another complex scalar $\tilde{b}$
with baryon charge  $Q_B(\tilde{b})=-2$. In the stringy description this scalar comes from ten-dimensional
three-form, see \cite{Louis} for a review, as well as Sec.~\ref{sec:tildebmass} and Appendix.

To conclude this section let us present the explicit metric of the singular conifold (with both $\beta$ and $b$ equal to zero), which will be used in the next section.
It has the  form  \cite{Candel}
\beq
ds^2_{6}=dr^2 + \frac{r^2}{6}(e_{\theta_1}^2+ e_{\varphi_1}^2 +e_{\theta_2}^2+ e_{\varphi_2}^2) +\frac{r^2}{9}e_{\psi}^2 ,
\label{conmet}
\eeq
where
\beqn
&& e_{\theta_1}= d\theta_1 , \qquad  e_{\varphi_1}= \sin{\theta_1}\, d\varphi_1\,,%\\[1mm]
\nonumber\\
&& e_{\theta_2}= d\theta_2 , \qquad  e_{\varphi_2}= \sin{\theta_2}\, d\varphi_2\,,%\\[1mm]
\nonumber\\
&& e_{\psi}= d\psi  + \cos{\theta_1}d\varphi_1+ \cos{\theta_2}d\varphi_2\,.
\label{angles}
\eeqn
Here $r$ is another  radial coordinate on the cone while the angles above are defined at $0\le \theta_{1,2}<\pi$,
$0\le \varphi_{1,2}<2\pi$, $0\le \psi<4\pi$.

\vspace{2mm}

The volume integral associated with this metric is
\beq
({\rm Vol})_{Y_6} = \frac{1}{108}\int r^5 \,dr \, d\psi\,  d\theta_1\,\sin{\theta_1} d\varphi_1 \,d\theta_2 \,
 \sin{\theta_2}d\varphi_2\,.
\label{Vol}
\eeq
The  radial coordinate, $\widetilde{r}$ defined in terms of matrix $w^{PK}$, see \eqref{tilder}  is related to
$r$ in (\ref{conmet}) via \cite{Candel}
\beq
r^2 = \frac32 \,\widetilde{r}^{4/3}\,.
\label{rtilder}
\eeq

\section {NS 3-form}
\label{sec:H_3}
\setcounter{equation}{0}

Now we switch on small quark masses $m_A$, $A=1,...,4$. We will use the effective 10D supergravity approach to
find the deformed background for our non-Abelian vortex string. As we already explained in the Introduction, instead of 
attempting to interpret  twisted mass terms in \eqref{wcp22} in terms of 10D gravity we take a different route.
We exploit the standard string theory approach of flux compactifications to 4D. Namely, we look for the solution of 10D gravity equations of motion with  non-zero NS 3-form $H_3$ and then interpret  $H_3$ in terms of quark masses. Note that the 
3-form  $H_3$ flux does not break  \ntwo supersymmetry in 4D \cite{Louis2}.

The motivation to consider the NS 3-form $H_3$ is as follows. We expect that quark masses should lift the Higgs
branch associated with the massless baryon $b$. In particular,  the mass term
\eqref{m_b} for the baryon $b$ appears for  generic values of quark masses as dictated for a BPS state with non-zero  baryonic charge \cite{ISY_b_baryon}.
On the other hand, generically, in type IIA compactifications a potential for K\"ahler moduli is generated by even-form RR fluxes,
 while the potential for complex structure moduli (like our $b$-modulus) is induced by the NS 3-form flux \cite{Louis2}.

As we already mentioned,  solving gravity equations of motion we will use the perturbation theory at small $H_3$. Namely, we will solve  equations of motion for $H_3$ using the  conifold metric neglecting the back reaction on the metric and the dilaton. These effects  appear in the quadratic order in $H_3$. We also assume that VEV of the baryon $b$ is large to make sure that the curvature of the conifold is everywhere small. This justify the  gravity approximation.

\subsection{Solution for NS 3-form at large $r$}

The bosonic part of the action of the type IIA supergravity in the Einstein frame is given by
\beqn
&& S_{10D} = %\frac1{2\kappa^2}\,
\int d^{10} x \sqrt{-G}\left\{ R - \frac12 G^{MN}\,\pt_M\Phi\,\pt_N\Phi 
\right.
\nonumber\\
&& -
\left.
\frac{e^{-\Phi}}{12}\,H_{MNL}H^{MNL}-\frac12\,e^{\frac{\Phi}{2}}\,F_4^2\right\} 
+ %\frac1{2\kappa^2}\,
T^2\int \frac12\, H_3\wedge C_3\wedge dC_3,
\label{10Daction}
\eeqn
where $G_{MN}$ and $\Phi$ are 10D metric and  dilaton, the string coupling $g_s=e^{\Phi}$. We also keep only NS 2-form $B_2$ with the field strength $H_3=dB_2$ and RR 3-form $C_3$ with
the field strength $F_4=dC_3$. We will need RR 3-form $C_3$ below to introduce the complex scalar $\tilde{b}$, the bosonic superpartner of $b$ in the 4D baryonic hypermultiplet. 
%The coupling $\kappa$ is determined by the tension of our non-Abelian vortex string \eqref{ten}, $\kappa^2 \sim 1/T^2$. 
%$\kappa^2 \sim g_s^2/T^4$ if we redefine the radial coordinate on the conifold to have the same dimension $(-1)$ as $x_{\mu}$, $r\to r/\sqrt{T}$.

In this section we will find a solution for the NS 3-form $H_3$ at large values of the radial coordinate $r$. Large means that  $\widetilde{r}$ is much larger then the minimal radius of the  $S_3$ of the deformed conifold, which is  equal to $\sqrt{|b|}$, see  \eqref{deformedconi}. This translates into $r\gg |b|^{1/3}$, see \eqref{rtilder}. At large $r$ we can neglect the deformation of the conifold and 
use the metric of the singular conifold. Thus, the  10D space has the structure $\mathbb{R}^4\times Y_6$ with the metric 
\beq
ds^2_{10} = T\,\left[-(dt)^2 + (d x^1)^2 + (d x^2)^2 + (d x^3)^2 \right] + \,ds^2_6,
\label{10met}
\eeq
where the metric of $Y_6$ is given by \eqref{conmet} and $T$ is the string tension \eqref{ten}.

For 2-form $B_2$ we use the ansatz introduced in \cite{KlebNekras,KlebTseytlin,KlebStrass,tseytlin} for type IIB ''compactifications''  on the conifold. Namely, we write
\beq
B_2= f_1(r)\, e_{\theta_1}\wedge e_{\varphi_1} + f_2(r)\, e_{\theta_2}\wedge e_{\varphi_2},
\label{B_2}
\eeq
where $f_1(r)$ and $f_2(r)$ are functions of the radial coordinate $r$, while angle differentials are defined in \eqref{angles}.
This gives for the 3-form field strength
\beq
H_3= f_1' \, dr\wedge e_{\theta_1}\wedge e_{\varphi_1} + f_2' \, dr \wedge e_{\theta_2}\wedge e_{\varphi_2},
\label{H_3}
\eeq
where prime denotes the derivative with respect to $r$. This 3-form is closed so the Bianchy identity is satisfied.

In order to find a non-zero solution for $H_3$ we have to either introduce a source like a D-brane or impose a non-trivial boundary condition for $H_3$. We use the latter option and specify the  boundary condition for $H_3$ at small $r$ in 
Sec. \ref{sec:H3_small_tau}.

The equation of motion for $H_3$ reads
\beq
d(e^{-\Phi}\ast H_3)= e^{-\Phi} d(\ast H_3) =0,
\label{H_3eqn}
\eeq
where $\ast $ denotes the  Hodge star and, as we explained above, we neglect the back reaction on the dilaton to the leading order in small $H_3$ and assumed that the dilaton is constant. 

The 10D-dual of $H_3$ is given by
\beq
\ast H_3 =  \frac{T^2}{3}\, dx^0 \wedge dx^1 \wedge dx^2 \wedge dx^3 \wedge e_{\psi} \wedge \left( e_{\theta_2}\wedge e_{\varphi_2}\, r f_1'
+ e_{\theta_1}\wedge e_{\varphi_1}\, r f_2'\right).
\label{astH_3}
\eeq
Substituting this into equation of motion \eqref{H_3eqn} we find three equations,
\beq
\pt_r(r\,f_1')=0, \qquad \pt_r(r\,f_2')=0
\label{feqs}
\eeq
and
\beq 
f_1'+ f_2' =0.
\label{f1+f2}
\eeq
The non-zero solution to these equations has the form 
\beq
f_1 = -f_2 = \mu_1 \log{r},
\label{f}
\eeq
where   $\mu_1$ is a small real parameter which we will interpret in terms of quark masses in Sec. \ref{sec:quarkmasses}.

This gives for the 3-form $H_3$
\beq
H_3=   \mu_1\,\frac{dr}{r}\wedge \left(e_{\theta_1}\wedge e_{\varphi_1} -   e_{\theta_2}\wedge e_{\varphi_2}\right).
\label{H_3solution}
\eeq
Essentially, this coincides with the solution obtained in \cite{KlebNekras,KlebTseytlin,KlebStrass} for NS 3-form in type IIB ''compactifications'' on the conifold.

\subsection{A generalization}

In this section we  make a generalization of our solution \eqref{H_3solution} for $H_3$. Let us define two real 3-forms
on $Y_6$,
\beq
\alpha_3\equiv \frac{dr}{r}\wedge \left(e_{\theta_1}\wedge e_{\varphi_1} -   e_{\theta_2}\wedge e_{\varphi_2}\right)
\label{alpha}
\eeq
and 
\beq
\beta_3\equiv e_{\psi}\wedge \left(e_{\theta_1}\wedge e_{\varphi_1} -   e_{\theta_2}\wedge e_{\varphi_2}\right)
\label{beta}
\eeq
They are both closed \cite{KlebNekras,KlebTseytlin},
\beq 
d\alpha_3 =0, \qquad d\beta_3 =0,
\eeq
Moreover, using the conifold metric \eqref{conmet} one can check that their 10D-duals are given by
\beqn
&& \ast\alpha_3 = -\frac{T^2}{3}\, dx^0 \wedge dx^1 \wedge dx^2 \wedge dx^3 \wedge\beta_3, 
\nonumber\\
&&\ast\beta_3 =  3\, T^2\, dx^0 \wedge dx^1 \wedge dx^2 \wedge dx^3 \wedge\alpha_3,
\label{alphabetadual}
\eeqn
where the first relation was already shown in the previous section, see \eqref{astH_3}. 

The above relations ensure that both 10D-dual forms satisfy equations of motion
\beq
d\ast\alpha_3 =0, \qquad d\ast\beta_3 =0.
\label{eqnofmotion}
\eeq
Therefore, we can generalize our solution \eqref{H_3solution} for $H_3$ writing
\beq
H_3 =\mu_1 \alpha_3 + \frac{\mu_2}{3} \,\beta_3,
\label{H_3sol}
\eeq
where $\mu_2$ is another real parameter, while the factor $\frac13$ is introduced for convenience. This $H_3$-form satisfy both Bianchi identity and equations of motion \eqref{H_3eqn}.

3-Forms \eqref{alpha} and \eqref{beta} form a basis similar to the simplectic basis of harmonic $\alpha$ and $\beta$ 3-forms for compact CYs, see for example review \cite{Louis}. In particular,
\beq
\int_{Y_6} \alpha_3 \wedge \alpha_3 = \int_{Y_6} \beta_3 \wedge \beta_3 = 0,
\label{aabb}
\eeq
while 
\beq
\int_{Y_6} \alpha_3 \wedge \beta_3 \sim -\int \frac{dr}{r} \sim - \log{\frac{R_{\rm IR}^3}{|b|}}. 
\label{ab}
\eeq
Here $R_{\rm IR}$ is the maximal value of the radial coordinate $r$ introduced to regularize the infrared logarithmic divergence,
while at small $r$ the integral is cut off by the minimal size of $S_3$ which is equal to $|b|$. This logarithmic 
behavior will play an important role below. Note that this logarithm is similar to the one, which determines the metric
for the $b$-baryon in \eqref{Sb} \footnote{Note that $R_{\rm IR}^3\sim \widetilde{R}_{\rm IR}^2$, see \eqref{rtilder}.}.

\subsection{Mass term for  $\tilde{b}$-baryon}
\label{sec:tildebmass}

Generically in type IIA compactifications the potential for complex structure moduli comes from two sources:
the  topological term  and the kinetic term for $B_2$ in the 10D action, namely, the last and the first terms in the second line in
\eqref{10Daction} respectively \cite{Louis2}. In this section we consider the topological term.

Following the standard procedure for compactifications on CY 3-folds \cite{Louis} we expand the 10D potential $C_3$ as
\beq
C_3= C_3^{4D} + \tilde{b}_1\,\alpha_3  + \frac13\,\tilde{b}_2\,\beta_3.
\label{C_3}
\eeq
Here $C_3^{4D}(x)$ is the part of the 10D potential $C_3$ oriented in 4D which depends only on 4D coordinates $x_{\mu}$.  Real 4D scalar fields $\tilde{b}_1$ and 
$\tilde{b}_2$ can be combined into a complex  scalar $\tilde{b}(x)=\tilde{b}_1(x) + i\tilde{b}_2(x)$ which together with the complex scalar $b$ form a bosonic part of the 4D baryonic hypermultiplet.

In Appendix we calculate  kinetic terms for 4D scalars $\tilde{b}_{1,2}$ using two last terms in \eqref{C_3} to calculate 
4-form field strength $F_4=dC_3$. 
 We show that scalars  $\tilde{b}_{1,2}$ acquire the same logarithmic metric as 
the complex scalar $b$, see \eqref{Sb} as dictated by \ntwo supersymmetry in 4D.

In order to obtain mass term for $\tilde{b}$ we following \cite{Louis2}  saturate the last factor $dC_3$ in the topological term in \eqref{10Daction} by the 4D potential 
$C_3^{4D}(x)$.   
Dualizing  it to a 4D scalar $e(x)$, $dC_3^{4D}(x)_{\mu\nu\delta\gamma}= e(x)\,\varepsilon_{\mu\nu\delta\gamma}$
we get
\beq
\int \frac12\, H_3\wedge C_3\wedge dC_3 = \frac12 \int d^4x \left\{ e \,\int_{Y_6} H_3\wedge C_3 \right\}.
\eeq
Now we substitute here our solution for $H_3$-form \eqref{H_3sol} and the expansion \eqref{C_3}. This gives
the contribution to the 4D action
\beq
  T^2\int d^4x \,e\,(\mu_2 \tilde{b}_1 - \mu_1 \tilde{b}_2)\,\log{\frac{R_{\rm IR}^3}{|b|}},
\eeq
where we used \eqref{aabb}, \eqref{ab}.

Combining this with the kinetic term for $C_3^{4D}$ in \eqref{10Daction} we get the part of the 4D action depending on the scalar $e$
\beq
S_{4D}^e = T^2\int d^4x \left\{\frac12 ({\rm Vol})_{Y_6}\,e^{\Phi/2}\, e^2 + e\,\left(\mu_2 \tilde{b}_1 - \mu_1 \tilde{b}_2\right)\,
\log{\frac{R_{\rm IR}^3}{|b|}}\right\}
\eeq
Here $({\rm Vol})_{Y_6} \sim R_{\rm IR}^6$ is the infinite volume of the conifold, see \eqref{Vol}. Integrating the scalar $e$ out we finally get
the mass term  for $\tilde{b}$
\beq
S_{mass}(\tilde{b}) =  -\frac{T^2\,e^{\Phi/2}}{2({\rm Vol})_{Y_6}}\,\int d^4x \,\left(\mu_2 
\tilde{b}_1 - \mu_1 \tilde{b}_2\right)^2\,
\left(\log{\frac{R_{\rm IR}^3}{|b|}}\right)^2|_{R_{\rm IR}\to\infty} \to 0
\label{massterm}
\eeq

We see that the mass term for the baryon $\tilde{b}$ vanishes. In contrast to type IIA compactifications on compact CY spaces, where the topological term in \eqref{10Daction} produces mass terms for scalar superpartners of complex structure moduli \cite{Louis2}, in our case it goes to zero   due to the non-normalizability of  $C_3^{4D}$.
In terms of quark masses we interpret the result in \eqref{massterm} as follows. NS 3-form $H_3$ corresponds to a particular choice of quark masses for which $m_b=0$, see \eqref{m_b}. We will come back to the interpretation of $H_3$ in terms of quark masses later in Sec.~\ref{sec:quarkmasses}. 

In fact this result is a test of  the consistency of  our approach. To see this note that the mass term \eqref{massterm} gives mass to a particular combination of $ \tilde{b}_1$ and $\tilde{b}_2$, while the orthogonal combination remains massless.  If this term were nonzero this would signal the breaking of  \ntwo supersymmetry in 4D. On the other hand, we know that quark masses do not break \ntwo supersymmetry in 4D SQCD.

\subsection{The potential for the baryon $b$}
\label{sec:pot}

As we already mentioned in the type IIA compactifications another source of the potential for complex structure moduli come from the kinetic term for $B_2$ in \eqref{10Daction}. Writing this term as 
\beq
- \frac{e^{-\Phi}}{12}\int H_3\wedge \ast H_3
\label{kinH_3}
\eeq
and substituting here our solution \eqref{H_3sol} we  get
\beq
 -T^2\,e^{-\Phi}\, (\mu_1^2 +\mu_2^2)\,\int d^4x \, \log{\frac{R_{\rm IR}^3}{|b|}},
\eeq
where we used \eqref{alphabetadual} and \eqref{ab}. Thus, the scalar potential has the form
\beq
V(b) = {\rm const}\,T^2\,e^{-\Phi}\, (\mu_1^2 +\mu_2^2)\, \log{\frac{R_{\rm IR}^3}{|b|}}
\label{pot}
\eeq

The  potential \eqref{pot} is proportional to the same infrared logarithm which enters  the metric \eqref{Sb} for the baryon 
$b$.
We see that the Higgs branch for $b$ is lifted and we have a run-away vacuum with VEV 
\beq
\bra|b|\ket \to R_{\rm IR}^3 \to \infty. 
\label{bVEV}
\eeq
In fact our solution \eqref{H_3sol} is found using the metric of the singular conifold and therefore is valid 
at $r\gg |b|^{1/3}$.
Thus, the potential \eqref{pot} can be trusted at small $|b|$ where the logarithm is large and   we cannot really use it at 
$|b|\sim R_{\rm IR}^3$. In the next section we consider the case of large $|b|$  and confirm our conclusion in \eqref{bVEV} that the VEV of the baryon $b$ tends to infinity.

\section {Deformed conifold}
\label{sec:H_3large_b}
\setcounter{equation}{0}

In the previous section we have seen that the 3-form $H_3$ lifts the Higgs branch for the baryon $b$ and pushes the VEV of $b$ towards large $|b|$, $|b|\sim R_{\rm IR}^3$. To study the behavior of the potential $V(b)$ in this case we have to find the solution
for $H_3$ on the deformed conifold assuming that the radial coordinate $r\sim |b|^{1/3}$. First, we briefly review the metric of the deformed conifold and then solve  equations of motion for $H_3$.

\subsection{Metric of the deformed conifold}

The metric of the deformed conifold  has the form \cite{Candel,Ohta,KlebStrass}
\beqn
ds_6^2 &=& \frac12\,|b|^{2/3}\,K(\tau)\left\{ \frac{1}{3K^3(\tau)}\left(d \tau^2 + e_{\psi}^2\right)
+ \cosh^2{\frac{\tau}{2}}\,\left(g_3^2+ g_4^2\right) 
\right.
\nonumber \\
&+& \left.
 \sinh^2{\frac{\tau}{2}}\,\left(g_1^2+ g_2^2\right)\right\},
\label{defconmet}
\eeqn
where angle differentials are defined as 
\beqn
&&
g_1= -\frac1{\sqrt{2}}\,(e_{\phi_1} + e_3), \qquad g_2= \frac1{\sqrt{2}}\,(e_{\theta_1} - e_4),
\nonumber\\
&&
g_3= -\frac1{\sqrt{2}}\,(e_{\phi_1} - e_3), \qquad g_4= \frac1{\sqrt{2}}\,(e_{\theta_1} + e_4),
\label{g_angles}
\eeqn
while 
\beq
e_3= \cos{\psi}\sin{\theta_2}\, d\varphi_2 - \sin{\psi}\,d\theta_2, \qquad 
e_4= \sin{\psi}\sin{\theta_2}\, d\varphi_2 + \cos{\psi}\,d\theta_2,
\eeq
see also \eqref{angles}.

Here
\beq
K(\tau)=\frac{(\sinh{2 \tau}-2 \tau)^{1/3}}{2^{1/3}\sinh{\tau}}
\eeq 
and the new radial coordinate $\tau$ is defined as
\beq
\widetilde{r}^2=|b|\,\cosh{\tau} = \left(\frac23\right)^{\frac32}\,r^3.
\label{tau}
\eeq
In the limit of large $\tau$ the metric \eqref{defconmet} reduces to the metric \eqref{conmet} of the singular conifold.

Results of the previous section show that we have a run-away vacuum with $|b|\sim R_{IR}^3$ so we are interested in the metric
\eqref{defconmet} in the limit of small $\tau$, $\tau\ll 1$. In this limit the metric of the deformed conifold takes the form
\beqn
ds_6^2|_{\tau\to 0} &=& \frac12\,|b|^{2/3}\left(\frac23\right)^{\frac13}\,\left\{ \frac12\,d \tau^2 + \frac12\,e_{\psi}^2
+ g_3^2+ g_4^2 
\right.
\nonumber \\
&+& \left.
 \frac{\tau^2}{4}\,\left(g_1^2+ g_2^2\right)\right\}.
\label{conmettau0}
\eeqn
The last term here corresponds to the collapsing sphere $S_2$, while the sphere $S_3$ associated with three  angular terms 
in the first line  has a fixed radius in the limit $\tau\to 0$ \cite{Candel,KlebStrass}. The radial coordinate $r$ approaches its minimal value with
\beq
r^3|_{min} = \left(\frac32\right)^{\frac32}\, |b|
\label{r_min}
\eeq
at $\tau=0$.

The square root of the determinant of the metric
\beq
\sqrt{g_6} \sim |b|^2\,\cosh^2{\frac{\tau}{2}}\sinh^2{\frac{\tau}{2}} |_{\tau\to 0} \sim |b|^2\, \tau^2
\label{det}
\eeq
vanishes at $\tau=0$, which  shows the degeneration of the conifold metric.

\subsection{NS 3-form at small $\tau$}
\label{sec:H3_small_tau}

In this section we consider an extrapolation of our solution for $H_3$ obtained for large $r$ to the region of small
values of $\tau$. For simplicity we will consider an extrapolation of the solution \eqref{H_3solution} taking $\mu_2=0$ in 
\eqref{H_3sol}.  

At large values of the radial coordinate $\tau$ the 2-form potential $B_2$ is proportional to
\beq
B_2 \sim e_{\theta_1}\wedge e_{\varphi_1} - e_{\theta_2}\wedge e_{\varphi_2}
\eeq
see \eqref{B_2} and \eqref{f}. The closed 2-form on the r.h.s  can be rewritten in terms of angular differentials $g$.
Explicitly,  we have 
\beq
e_{\theta_1}\wedge e_{\varphi_1} - e_{\theta_2}\wedge e_{\varphi_2}= g_1\wedge g_2 + g_3\wedge g_4.
\eeq
Therefore, it is natural to look for a solution for $B_2$ at arbitrary $\tau$ using the ansatz,
\beq
B_2 = p(\tau)\,  g_1\wedge g_2 + k(\tau)\, g_3\wedge g_4,
\label{B_2ansatz}
\eeq
where $p$ and $k$ are functions of the radial coordinate $\tau$. This ansatz was used in \cite{KlebStrass} for 
the type IIB ''compactification'' on the deformed conifold.

The ansatz \eqref{B_2ansatz} gives for the field strength 
\beq
H_3 = p'\, d\tau\wedge g_1\wedge g_2 + k'\, d\tau\wedge g_3\wedge g_4 - \frac12\,(p-k)\, e_{\psi}\wedge 
(g_1\wedge g_3 + g_2\wedge g_4),
\label{H_3tau}
\eeq
where we used the identity \cite{KlebStrass}
\beq
d(g_1\wedge g_2 - g_3\wedge g_4) = - e_{\psi}\wedge (g_1\wedge g_3 + g_2\wedge g_4).
\label{id1}
\eeq
Here primes denote derivatives with respect to $\tau$.

In particular, at large $\tau$ we have $p=k=\mu_1/3$, where we rewrite solution \eqref{H_3solution} in terms of $\tau$
using relation \eqref{tau} at $\tau\to\infty$.

Calculating the 10D-dual of \eqref{H_3tau} we get 
\beqn
\ast H_3 &=&  T^2\, dx^0 \wedge dx^1 \wedge dx^2 \wedge dx^3 \wedge \left\{p'\, \coth^2{\left(\frac{\tau}{2}\right)}\,e_{\psi}\wedge g_3\wedge g_4   
\right.
\nonumber\\
&+&
 k'\, \tanh^2{\left(\frac{\tau}{2}\right)}\,e_{\psi}\wedge g_1\wedge g_2
\nonumber\\
&-&
\left.
\frac12\,(p-k)\, d\tau\wedge (g_1\wedge g_3 + g_2\wedge g_4) \right\}. 
\label{astH_3tau}
\eeqn
Then the equation of motion \eqref{H_3eqn} gives two equations
\beqn
&& 4\pt_{\tau}\left( \frac{p'}{\tau^2}\right) - \frac12\,(p-k) =0,
\nonumber\\
&& \frac12\pt_{\tau}\left( \tau^2 k'\right) + (p-k) =0,
\label{pkeqs}
\eeqn
which we write down in the limit of small $\tau$. To derive \eqref{pkeqs} we used the identity \cite{KlebStrass}
\beq
d(g_1\wedge g_3 + g_2\wedge g_4) =  e_{\psi}\wedge (g_1\wedge g_2 - g_3\wedge g_4).
\label{id2}
\eeq

The solution of these equations at small $\tau$ has the form
\beq
k(\tau) \approx \mu_1\,\tau, \qquad p(\tau) \approx - \frac{\mu_1}{80}\,\tau^5,
\label{pk}
\eeq
up to an overall constant.
%Solutions for $H_3$ obtained in \cite{KlebStrass} for the case of type IIB ''compactification'' have a similar, but not identical behavior at small $\tau$.  

To conclude this section, we note that at $\tau=0$ our solution \eqref{H_3tau} tends to a constant
\beq
H_3(\tau=0) = \mu_1\,d\tau\wedge g_3\wedge g_4,
\label{bc}
\eeq
which we impose as  boundary conditions at $S_3$, which does not shrinks at $\tau=0$.  These boundary conditions ensure a non-zero solution for $H_3$.

\subsection{The potential for the baryon $b$ at large $|b|$}

Substituting solutions for $H_3$ and its 10D-dual given by \eqref{H_3tau} and \eqref{astH_3tau} to the kinetic term \eqref{kinH_3}
we get 
\beq
- \frac{e^{-\Phi}}{12}\int H_3\wedge \ast H_3 \sim -\mu_1^2 \,T^2\, e^{-\Phi}\,\int d^4 x \,d\tau \,\tau^2,
\eeq
where only the function $k(\tau)$ \eqref{pk} contributes to the leading order in $\tau$.
Thus, the potential for $b$ takes the form 
\beq
V(b) = {\rm const}\, \mu_1^2\, T^2\, e^{-\Phi}\, \tau_{{\rm max}}^3,
\label{tau3}
\eeq
where $\tau_{{\rm max}}$ is the infrared cutoff with respect to the radial coordinate $\tau$ related to $R_{\rm IR}$ as follows
\beq
|b|\cosh(\tau_{{\rm max}}) = \left(\frac23\right)^{\frac32}\,R_{\rm IR}^3,
\eeq
see \eqref{tau}.

As we already explained,  we expect that in our run-away vacuum $b$ is large, close to $R_{\rm IR}$, therefore
$\tau_{{\rm max}}$ is small. Expanding $\cosh{\tau}$ at small $\tau$ we get
\beq
\tau_{{\rm max}} \sim \sqrt{\frac{R_{\rm IR}^3- \left(\frac32\right)^{\frac32}|b|}{|b|}}.
\eeq
This gives  the potential for the baryon $b$ at large $|b|$
\beq
V(b) = {\rm const}\, \mu_1^2\, T^2 e^{-\Phi}\, \left[\frac{R_{\rm IR}^3- \left(\frac32\right)^{\frac32}|b|}{|b|}\right]^{\frac32}.
\label{potlargeb}
\eeq
We see that to minimize the potential $|b|$ becomes large and approaches the infrared cutoff,
\beq
\bra|b|\ket = \left(\frac23\right)^{\frac32}\,R_{\rm IR}^3 \to \infty.
\label{VEVb}
\eeq
As we expected earlier in Sec.~\ref{sec:pot}, we get  a run-away vacuum. 

In fact, $\tau_{{\rm max}}^3$
which enters \eqref{tau3} is the volume of the  three dimensional ball bounded by the  sphere $S_2$ of the conifold with the maximum radius $\tau_{{\rm max}}$. It shrinks to zero as $b$ tends to its VEV \eqref{VEVb}.
To avoid singularities we can regularize the size of $S_2$  introducing small non-zero $\beta$, which  makes the conifold ''slightly resolved'' , see \eqref{D-term}. We  take the limit $\beta\to 0$ at the last step. Then the value of the potential and all its derivatives vanish in the vacuum \eqref{VEVb} at $|b|= \bra|b|\ket$, for example 
\beq
V(b)|_{|b|= \bra|b|\ket} = {\rm const}\, \mu_1^2 \,T^2 e^{-\Phi}\, \frac{\beta^3}{R_{\rm IR}^{9/2}} \to 0.
\label{VatVEV}
\eeq
In particular, the mass term for $b$ is zero. This is  in accordance with  \ntwo supersymmetry in 4D SQCD, since  the  mass
 of  $\tilde{b}$ is zero, see \eqref{massterm}. 

To summarize, the $H_3$-form flux produces following effects.
\begin{itemize}
\item The  Higgs branch of the baryon $b$ in 4D SQCD is lifted. 
\item The vacuum is of a run-away type $\bra|b|\ket \to\infty$.
\item At the run-away vacuum  the sphere $S_2$  of the conifold degenerates, while the radius of the sphere $S_3$ tends to infinity.
\end{itemize}
 We will interpret this degeneration in terms of \ntwo SQCD in Sec. \ref{sec:degeneration}.

\section {Interpretation in terms of 4D SQCD }
\label{sec:quarkmasses}
\setcounter{equation}{0}

\subsection{3-form flux in terms of quark masses}

In this section we interpret the NS 3-form $H_3$ in terms of quark masses of 4D \ntwo SQCD. For $N_f=4$ we have four complex mass parameters. However, a shift of the complex scalar $a$, a superpartner of  the U(1) gauge field, produces an overall shift of  quark masses. This can be also seen in the \wcpt model \eqref{wcp22} on the world sheet of the non-Abelian string; a constant shift of the scalar  $\sigma$ makes an overall shift of quark masses. Thus, in fact we have three independent complex mass parameters in our 4D SQCD. For example, we can choose  three mass differences 
\beq
m_1-m_2, \qquad m_3-m_4, \qquad m_1-m_3
\label{massdifferences}
\eeq 
as independent parameters.

On the string theory side our solution \eqref{H_3sol} for the 3-form $H_3$ is parametrized by two real parameters $\mu_1$ and 
$\mu_2$.
Thus, we expect that non-zero $H_3$-flux can be interpreted in terms of a particular choice of quark masses, subject to  two complex constraints.

One constraint we already discussed. We know that $H_3$ does not produces a mass term for the  $b$-baryon. This ensures that
\beq
m_1 +m_2 -m_3 -m_4 =0,
\label{constraint1}
\eeq
see \eqref{m_b}.

To derive the second constraint we consider the exact twisted superpotential for the \wcpt model \eqref{wcp22}
  obtained by integrating out $n$ and $\rho$ supermultiplets. It is a generalization \cite{HaHo,DoHoTo}
of the CP($N-1$) model superpotential \cite{AdDVecSal,ChVa,W93,Dorey} of the  Veneziano-Yankielowicz  type \cite{VYan}.
In the present case $N_f = 2N = 4$ it reads:
\begin{multline}
	 {\cal W}_{\rm WCP}(\sigma)= \frac{1}{4\pi}\Bigg\{ 
	 	\sum_{P=1,2} \left( \sqrt{2} \, \sigma + m_P \right) \ln\left( \sqrt{2} \, \sigma + m_P \right)
	 	\\
	 	- \sum_{K=3,4} \left( \sqrt{2} \, \sigma + m_K \right) \ln\left( \sqrt{2} \, \sigma + m_K \right)
	 	+ 2 \pi \,  \sqrt{2} \, \sigma  \, \beta
	 	+ \text{const}
	 \Bigg\}\,,
\label{WCPsup}
\end{multline}
where we use one and the same notation $\sigma$ for the  twisted superfield \cite{W93} and its lowest scalar
component. 

To study the vacuum structure of the theory we minimize this superpotential with respect to $\sigma$ to obtain the 2D vacuum equation
\begin{equation}
	\prod_{P=1,2}\left(\sqrt{2} \, \sigma + m_P \right) 
		= e^{- 2 \pi \beta} \, \prod_{K = 3,4} \left(\sqrt{2} \, \sigma + m_K \right) \,.
\label{2d_equation}	
\end{equation}
Consider the limit $\beta\to 0$ in this equation. In this limit \eqref{2d_equation} reduces to the following quadratic equation
\beq
2\pi\beta\,(\sqrt{2}\,\sigma)^2 + (\sqrt{2}\,\sigma)(m_1 +m_2 -m_3 -m_4) + m_1 m_2-m_3 m_4 =0.
\label{quad_eqn}
\eeq
The second term here is zero due to the constraint \eqref{constraint1}. Two roots of \eqref{quad_eqn} read
\beq
\sqrt{2}\,\sigma = \pm \left[ \frac1{2\pi\beta}\, ( m_1 m_2-m_3 m_4)\right]^{1/2}|_{\beta\to 0} \to \infty
\eeq

However,  infinite VEVs of $\sigma$ cost an infinite energy for non-zero $H_3$. To see this, observe that from the string theory side we learned that  the VEV of $b$ goes to infinity, see \eqref{VEVb}. Then eq. \eqref{tau} shows that the radial coordinate $\widetilde{r}$ is very large, which means that (some of)  $n$'s and $\rho$'s become  infinitely large, see \eqref{tilder}. Given the infinite VEV of $\sigma$ this makes first two terms in the second line of the  \wcpt action \eqref{wcp22} infinite.

 To avoid this we impose  the second constraint, namely
\beq
m_1 m_2-m_3 m_4 =0.
\label{constraint2}
\eeq
 Now, let us solve two constraints in  \eqref{constraint1} and \eqref{constraint2} for masses $m_3$ and $m_4$ as functions of 
$m_1$ and $m_2$. Finding, say, $m_4$ from \eqref{constraint1} and substituting it into \eqref{constraint2} we get a quadratic 
equation for $m_3$,
\beq
m_3^2 -m_3(m_1+m_2) + m_1m_2 =0.
\eeq
Two roots of this equation read
\beq
m_3=m_1, \qquad m_4=m_2
\label{option1}
\eeq
and 
\beq
m_3=m_2, \qquad m_4=m_1.
\label{option2}
\eeq
These two options are essentially the same, up to permutation of quarks $q^3$ and $q^4$. Let us choose the first option in 
\eqref{option1}.

%Substituting \eqref{option1} into vacuum equation \eqref{2d_equation} we see that at $\beta=0$ this equation is satisfied for any $\sigma$. This means that for this choice of quark masses at $\beta=0$ a Coulomb branch opens up and sigma can take arbitrary values.

The above arguments lead us to the conclusion that the $H_3$-flux can be interpreted in terms of the  single mass difference
$(m_1-m_2)$. Thus, we complexify parameters $\mu_{1,2} $ and  identify 
\beq 
\mu \equiv \mu_1 + i\mu_2 = \sqrt{\frac{g_s}{T}}\, (m_1-m_2), \qquad m_3=m_1, \qquad m_4=m_2.
\label{mum}
\eeq

Now, in terms of quark masses our solution \eqref{H_3sol} for the NS 3-form reads
\beq
H_3 = \sqrt{\frac{g_s}{T}}\,\left[{\rm Re} (m_1-m_2)\,\alpha_3 + \frac13\,{\rm Im} (m_1-m_2)\,\beta_3\right],
\label{H_3m}
\eeq
while the potential \eqref{pot} for the baryon $b$ becomes
\beq
V(b) = {\rm const}\,T\,|m_1-m_2|^2\, \log{\frac{R_{\rm IR}^3}{|b|}}.
\label{potm}
\eeq
Similarly we can rewrite our solution \eqref{H_3tau} for $H_3$ obtained at large $|b|$ in terms of $(m_1-m_2)$.

\subsection{Degeneration of the conifold}
\label{sec:degeneration}

If we take the limit of large $|m_1-m_2|\gg \sqrt{\xi}$ in 4D SQCD keeping the constraint \eqref{option1}  non-Abelian degrees of freedom decouple and U(2) gauge theory
flows to \ntwo supersymmetric Abelian theory with the gauge group U(1)$\times$U(1) and $N_f=4$ quark flavors. For example, 
off-diagonal gauge fields acquire large masses $\sim |m_1-m_2|$ \footnote{In addition to masses $m_G\sim g\sqrt{\xi}$ due to the Higgs mechanism, see \cite{SYrev} for a review.}  and decouple. However, in this paper 
we consider limit of small quark masses on the  string theory side and therefore, cannot make $|m_1-m_2|$  large. At small
 $|m_1-m_2|\ll\sqrt{\xi}$ our 4D SQCD remains to be U(2) gauge theory.

On the other hand in the world sheet \wcpt model  on the non-Abelian string the story is different. This theory is conformal and has no scale. Therefore, once we switch on non-zero $(m_1-m_2)$ there is no notion of smallness of this deformation. We see a dramatic effect as soon as $(m_1-m_2)$ becomes non-zero.

Classically there are two branches in \wcpt model \eqref{wcp22} associated with VEVs of $\sigma$
\beq
\sqrt{2}\,\sigma= -m_1
\label{sigma1}
\eeq
or 
\beq
\sqrt{2}\,\sigma= -m_2.
\label{sigma2}
\eeq
Let us choose the first branch for definiteness. Then fields $n^{P=2}$ and $\rho^{K=4}$ acquire masses $|m_1-m_2|$ and decouple
in the infrared. The low energy effective theory at scales below $|m_1-m_2|$ is the \wcpo model
\begin{equation}
\begin{aligned}
	&S = \int d^2 x \left\{
	\left|\nabla_{\alpha} n^{1}\right|^2 
	+\left|\tilde{\nabla}_{\alpha} \rho^3\right|^2
	+\frac1{4e^2}F^2_{\alpha\beta} + \frac1{e^2}\,
	\left|\pt_{\alpha}\sigma\right|^2
	\right.
	\\[3mm]
	&+\left.
	2\left|\sigma\right|^2 \left|n^{1}\right|^2 
	+ 2\left|\sigma\right|^2\left|\rho^3\right|^2
	+ \frac{e^2}{2} \left(|n^{1}|^2-|\rho^3|^2 - {\rm Re}\,\beta \right)^2
	\right\},
	\end{aligned}
\label{wcp11}
\end{equation}
where much in the same way as in \eqref{wcp22}, the limit $e^2\to\infty$ is assumed and we made a shift$(\sqrt{2}\sigma +m_1) \to \sqrt{2}\sigma$.

The number of real degrees of freedom in \eqref{wcp11} is $4-1-1=2$ where 4 is the number of real degrees of freedom of $n^1$ and $\rho^3$ and much in the same way as in \eqref{wcp22}, we subtract 2 due to the $D$-term constraint in \eqref{wcp11} and 
the U(1) phase eaten by the Higgs mechanism.

Physically \wcpo model describes an  Abelian semilocal vortex string supported in \ntwo supersymmetric U(1) gauge theory with 
$N_f=2$ quark flavors.  This vortex has no orientational moduli, but it  has one complex size modulus $\rho^3$, see 
\cite{AchVas,SYsem,Jsem}.  Thus, we can think that upon switching on $(m_1-m_2)$ a non-Abelian string flows to an Abelian one.

The low energy \wcpo model is also conformal.  Moreover,  it was shown in \cite{AharSeib} that in the non-linear sigma model formulation it flows to a free theory on $\mathbb{R}^2$ in the infrared.
Thus, in fact, switching on $(m_1-m_2)$ with constraint  \eqref{option1} does not break the conformal invariance on the world sheet. It just reduces the number of degrees of freedom  transforming a non-Abelian string into an Abelian one. The string theory which one would  associate with the sigma model \eqref{wcp11} is non-critical. 

This supports our interpretation of the  flux ''compactification'' on the conifold in terms of quark masses.
On the string theory side switching on $(m_1-m_2)$ is reflected in the degeneration of the conifold, which effectively reduces its dimension. Also in the limit 
$|b|\to\infty$ the radius of the sphere $S_3$ of the conifold  becomes infinite and it tends to a flat three dimensional space. This matches the field theory result \cite{AharSeib} that \wcpo model flows to a free theory in the infrared. It would be tempting to interpret the extra coordinate of the sphere $S_3$ of the conifold in the limit $|b|\to\infty$ as a Liouville coordinate for a non-critical string associated with the sigma model \eqref{wcp11}. This is left for a future work.

To conclude, we note that although  4D SQCD gets  just slightly deformed as we switch on small $(m_1-m_2)$ at weak coupling, this deformation becomes much more pronounced at strong coupling. Namely, at $\tau_{SW}=1$ (which corresponds to 
$\beta=0$ in \eqref{wcp22}) the non-perturbative Higgs branch gets lifted and we have a run-away vacuum \eqref{VEVb} for the 
 stringy baryon $b$.

\section{Conclusions}
\label{sec:conclusions}
\setcounter{equation}{0}

In this paper we considered the NS 3-form flux ''compactification'' for the string theory of  the critical non-Abelian vortex  supported in \ntwo SQCD with gauge group U(2) and $N_f=4$ quark flavors. Using supergravity approach we found a solution for the 3-form $H_3$ to the leading order at small $H_3$ neglecting the back reaction on the conifold metric and the dilaton. The non-zero 3-form $H_3$ generates a potential  for the complex structure modulus $b$ of the conifold, which is interpreted as a BPS baryonic hypermultiplet in 4D SQCD at strong coupling.
This potential lifts the Higgs branch formed by VEVs of $b$ and leads to a run-away for $b$, $\bra |b|\ket\to\infty$.

We interpret the 3-form $H_3$ as a quark mass deformation of 4D SQCD. The reason for this interpretation is that there is no
other deformation in 4D SQCD preserving \ntwo supersymmetry, which produces  a scalar potential. We use  field theory arguments to relate the 
3-form $H_3$ to the quark mass difference $(m_1-m_2)$, see \eqref{H_3m}, subject to the constraint \eqref{option1}.

At the run away vacuum the conifold degenerates to  lower dimensions. This qualitatively matches with a flow  to the \wcpo model \eqref{wcp11} on the string world sheet, expected if one  switches on the mass difference   $(m_1-m_2)$. This  flow can be thought of as a flow from a non-Abelian string to an Abelian one.

As a direction of a future research we can mention the finding of exact  solutions of 10D supergravity equations with non-zero
NS 3-form flux. This would allow us to describe the limit of large $(m_1-m_2)$ in terms of the string theory. In this limit
non-Abelian degrees of freedom in 4D SQCD with U(2) gauge group decouple and the theory flows to an Abelian U(1)$\times$U(1)
theory.

Also a challenging problem is to find a supergravity deformations associated with other choices of quark masses in 4D SQCD.

\section*{Acknowledgments}

The author is grateful to A. Gorsky,  E. Ievlev, A. Losev and M. Shifman  for very useful and 
stimulating discussions. This work  was  supported by William I. Fine Theoretical Physics Institute  of the  University 
of Minnesota.

\section*{Appendix. \\Kinetic term for $\tilde{b}$-baryon}

In this Appendix we calculate  kinetic terms for 4D scalars $\tilde{b}_{1,2}$ which arise from the expansion of the 
10D 3-form potential  $C_3$ in \eqref{C_3}. Assuming that scalars $\tilde{b}_{1,2}$ depend only on 4D coordinates $x_{\mu}$ 
and dropping the 4D 3-form $C_3^{4D}$ (we consider it in Sec. \ref{sec:tildebmass}) we write
\beq 
F_4=d C_3= \pt_{\mu} \tilde{b}_{1}\, dx^{\mu}\wedge \alpha_3 + \frac13\,\pt_{\mu} \tilde{b}_{2}\,  dx^{\mu}\wedge\beta_3.
\eeq
Calculating the 10D dual of this form we get
\beqn
&&\ast F_4= -\frac{1}{3\,3!}T^2\,\varepsilon_{\mu\nu\gamma\delta}\, \pt^{\mu}\tilde{b}_1\, dx^{\nu}\wedge dx^{\gamma}\wedge dx^{\delta}\wedge \beta_3 
\nonumber\\
&+& \frac{1}{3!}\,T^2\varepsilon_{\mu\nu\gamma\delta}\, \pt^{\mu}\tilde{b}_2\, dx^{\nu}\wedge dx^{\gamma}\wedge dx^{\delta}\wedge \alpha_3,
\label{astF_3}
\eeqn
where we used \eqref{alphabetadual}

Now writing the kinetic term for $C_3$ in \eqref{10Daction} as 
\beq
\int F_4\wedge\ast F_4
\eeq
we finally get
\beq
S_{{\rm kin}}(\tilde{b}) = T\int d^4x \left\{|\pt_{\mu} \tilde{b}_1|^2 + |\pt_{\mu} \tilde{b}_2|^2\right\}\,
\log{\frac{R_{\rm IR}^3}{|b|}}\,,
\label{Stildeb}
\eeq
where we used \eqref{aabb} and \eqref{ab}.

We see that scalars $\tilde{b}_{1,2}$ have the same logarithmic metric as the complex scalar $b$. As we already mentioned, the complexified scalar $\tilde{b}$ together with $b$ form a bosonic part of the 4D baryonic hypermultiplet.

\renewcommand{\theequation}{A.\arabic{equation}}
\setcounter{equation}{0}

\end{document}